\documentclass[twocolumn,prl,aps,superscriptaddress]{revtex4}
\draft
\usepackage{amsmath}
\usepackage{graphicx}
\begin{document}
\title{Surface-Enhanced Plasmon Splitting in a
Liquid-Crystal-Coated Gold Nanoparticle}
\author{Sung Yong Park}
\affiliation{Department of Physics, Ohio State University, Columbus, OH 43210}
\affiliation{Chemistry Department, Northwestern University, Evanston, IL 60208}
\author{D. Stroud}
\affiliation{Department of Physics, Ohio State University, Columbus, OH 43210}
\begin{abstract}
We show that, when a gold nanoparticle is coated by a thin layer of
nematic liquid crystal, the deformation produced by the nanoparticle
surface can enhance the splitting of the nanoparticle surface
plasmon. We consider three plausible liquid crystal director
configurations in zero electric field: boojum pair (north-south pole
configuration), baseball (tetrahedral), and homogeneous. From a
calculation using the Discrete Dipole Approximation, we find that
the surface plasmon splitting is largest for the boojum pair,
intermediate for the homogeneous, and smallest for the baseball
configuration. The boojum pair results are in good agreement with
experiment.  We conclude that the nanoparticle surface has a strong
effect on the director orientation, but, surprisingly, that this
deformation can actually enhance the surface plasmon splitting.
\end{abstract}
\maketitle

Colloidal suspensions and emulsions in anisotropic host media, or
coated with anisotropic media such as liquid crystals have attracted
considerable recent interest, since these systems exhibit a novel
class of colloidal interactions~\cite{poulin} or the possibility of
forming tetravalent binding sites of colloids~\cite{nelson}. The
alignment of the principal axis of anisotropic media (known as the
director) around a colloidal surface is a important feature; this
new colloidal interaction comes from homeotropic alignment,
perpendicular to the surface, while tetravalent binding sites from
tangential alignment to the surface.

These systems have also been studied for optical
applications~\cite{busch, expt, muller, ps}. For example, related
to the emerging nanotechnology, a recent experiment has
demonstrated nanoscale control of optical properties on a length
scale much less than a light wavelength~\cite{muller}. The authors
observed the surface plasmon splitting of a gold nanoparticle
coated with a thin layer of nematic liquid crystal: the
surface plasmon resonance
frequency in the scattering cross section was shown to depend on
the angle between the director and the polarization of incident
light. They also showed that, by rotating the director using a
static electric field, they could control the change in surface plasmon
frequency.

A recent calculation~\cite{ps} showed that this method could
indeed produce a measurable change in the surface plasmon frequency, assuming
that the director was oriented in the same direction everywhere
within the coating or the host media. But in reality, the director
field should be influenced near the metal surface~\cite{lubensky},
and thus the alignment of the liquid crystal director field near the
nanoparticle surface may reduce the surface plasmon splitting to hinder some
possible applications of this system which involve electrical
control of its optical properties. Thus, it is important to
investigate quantitatively how an inhomogeneous director field may
influence the optical properties of these coated metal particles.

In this Letter,
we report on novel optical phenomena arising from the surface deformation
(or alignment) of the liquid crystal director field.
Our main finding is that this deformation
can actually {\em enhance} this splitting of the surface plasmon
frequency.   We consider three plausible morphologies for this
director field, and, using the Discrete Dipole
Approximation (DDA)~\cite{dda,dda1}, compute the deformation-induced
splitting of the surface plasmon frequency into two  different
frequencies polarized parallel and perpendicular to the director.
For the expected ``boojum pair" configuration, the splitting is
enhanced, relative to that of a uniform director field, and agrees
very well with the measured splitting~\cite{muller}.
Potential importance of this work will be discussed below.

First, we discuss several possible morphologies for the liquid crystal
director field in an nematic liquid crystal layer on a gold nanoparticle.
The actual
morphology will depend strongly on the exact experimental setup.
We simply assume that, as illustrated in Fig. 1(a), a spherical
coated gold nanoparticle in an air host sits on the substrate
which is coated with a thin alignment layer, so that near the
substrate the director will be aligned in the $x$ direction. We
also assume that the liquid crystal director is aligned parallel to the
surface at both interfaces. This assumption is reasonable: for a
metal-liquid crystal boundary, a suitable surface treatment can modify the
boundary condition to the tangential boundary condition (director
parallel to the surface)~\cite{poulin,experimental}. Also if the
coating is thin, we can assume a tangential boundary condition at
the liquid crystal-air interface~\cite{experimental1}. Since the anchoring
force at the nanoparticle surface can be modified by suitable
surface preparation, the director may sometimes satisfy a more
general boundary condition, which will not, however, be considered
in this Letter.

The standard analytical description of liquid crystal morphology is based on the
Frank elastic free energy~\cite{lc}.  If there is no external
electric field, assuming the tangential boundary condition
there are two likely morphologies for a
spherical nematic liquid crystal layer on a spherical nanoparticle.
For a relatively
thick layer, the director field is expected to form a north-south
pole type (or boojum pair) morphology. But if the nematic liquid crystal
layer is very
thin, so that it can be treated like a two-dimensional surface, the
director field is expected to take a baseball-like
morphology~\cite{nelson}.
These two possibilities are shown in Fig.\ 1.   We will calculate
the scattering cross section for an liquid-crystal-coated gold nanoparticle,
considering both possibilities.

In our actual optical calculation, we have simplified these two
morphologies for computational convenience.  For the boojum pair
morphology, we assume that the director field is everywhere
parallel to the surface and points along a latitude between north
and south pole. This assumption is reasonable because of
anchoring, which forces the director field to be everywhere
parallel to the particle surface. The liquid crystal director orientation is
also affected by the alignment layer in the substrate. If we
simply assume a tangential boundary condition where the liquid-crystal-coated
nanoparticle touches the substrate, this condition limits the pole
positions. For example, in the boojum pair configuration, the
alignment layer forces the line between north and south poles to
be oriented in the $x$ direction, i.\ e.\ parallel to the substrate.

For the baseball morphology, there are four poles, at the vertices
of a tetrahedron. Because of the substrate-imposed anchoring, we
assume that two of these are located in lower hemisphere, and that
the line between these two poles is parallel to the $x$ axis,
while the other two are located in
upper hemisphere, joined by a line parallel to the $y$ axis. To use
the DDA, we further simplify this
configuration by assuming that the director fields in the lower and
upper hemispheres are both homogeneous, and point in the $x$ and $y$
directions, i.\ e.\ parallel to the substrate surface in both cases.

To calculate the scattering cross section of the
liquid-crystal-coated nanoparticle, we use an approach known as the
DDA~\cite{dda,dda1}. Given the director
fields of the liquid crystal coating, we calculate the scattering
cross section $C_{\rm scat}$ of a single nanoparticle
as if none of the other particles
were present.  This amounts to neglecting multiple scattering
corrections among the different nanoparticles, consistent with the
experiment of Ref.\ \cite{muller}. Also, we ignore the contribution
of the substrate to the scattering cross section (although we still
include its effects in aligning the liquid crystal director around
the nanoparticle as described above). Thus, $C_{\rm scat}$ can be
calculated without an orientational average over the scattering
object.

As originally formulated, the DDA allows
calculation of properties such as the extinction or scattering
coefficient of an irregularly shaped object having a complex,
frequency-dependent dielectric constant $\epsilon(\omega)$, embedded
in a homogeneous medium of real dielectric constant $\epsilon_h$. In
this approximation, the object is taken as an array
of point dipoles $(i= 1,\ldots, N)$ with dipole moments ${\bf P}_i$
and polarizabilities $\alpha_i$, located at positions ${\bf r}_i$.
The extension of the DDA to treat an
irregularly shaped object which has an anisotropic and spatially
inhomogeneous dielectric tensor can be easily done by introducing a
position-dependent tensor polarizability $\tilde{\alpha}_i$, and
thus the DDA can be applied to our case of
an liquid-crystal-coated nanoparticle with a spatially varying
director field. Thus ${\bf P}_i$ is expressed as
\begin{equation}
{\bf P}_i = \tilde{\alpha}_i {\bf E}_i,\label{eq:2}
\end{equation}
where
${\bf E}_i$, the electric
field at ${\bf r}_i$, is due to the incident wave ${\bf E}_{{\rm
inc},i}={\bf E}_0\exp({\bf k} \cdot{\bf r}_i - \omega t)$, plus the
contribution of each of the other $N-1$ dipoles:
\begin{equation}
\label{eq:3} {\bf E}_i = {\bf E}_{{\rm inc},i} - \sum_{j \neq i}{\bf
A}_{ij}\cdot{\bf P}_j.
\end{equation}

In the DDA, the product ${\bf
A}_{ij}\cdot{\bf P}_j$ can be expressed as
\begin{eqnarray}
\label{eq:4} {\bf A}_{ij}\cdot{\bf P}_j &= &
\frac{e^{ikr_{ij}-i\omega t}}{r_{ij}^3} \{k^2{\bf
r}_{ij}\times({\bf r}_{ij} \times {\bf P}_j)
 \nonumber \\
& + & \frac{1-ikr_{ij}}{r_{ij}^2}[r_{ij}^2{\bf P}_j -3{\bf r}_{ij}
({\bf r}_{ij}\cdot {\bf P}_j)]\}.
\end{eqnarray}
Here ${\bf r}_{ij} = {\bf r}_i - {\bf r}_j$ and $k = \omega/c \equiv
2\pi/\lambda$, $c$ and $\lambda$ being being the speed of light and
wavelength in vacuum.  Eqs.\ (\ref{eq:2})-(\ref{eq:4}) form a
coupled set of $3N$ equations, which can be solved for the $N$
dipole moments ${\bf P}_i$ using the complex-conjugate gradient
method combined with fast Fourier transforms~\cite{dda}.
Given the ${\bf P}_i$'s, $C_{\rm scat}$
for a given coated particle is obtained from the relation
\begin{equation}
C_{\rm scat} = \frac{4\pi k}{|{\bf E}_0|^2}\sum_{i = 1}^{N} \left[
\frac{2}{3}k^3|{\bf P}_i|^2 -{\rm Im}\left({\bf P}_i\cdot{\bf E}_{{\rm
self},i}^*\right) \right],
 \label{eq:5}
\end{equation}
where ${\bf E}_{{\rm self},i}={\bf E}_i-{\bf E}_{{\rm inc},i}$.

Now we discuss the relation between the local polarizability
tensor $\tilde{\alpha}$ of these point objects and the local
dielectric tensor $\tilde{\epsilon}(\omega)$. To carry out the
above calculation, one needs an expression for the polarizability
tensor $\tilde{\alpha}$. For a material with an {\em isotropic}
scalar dielectric function $\epsilon(\omega)$, one simple means of
connecting a scalar polarizability $\alpha$ to $\epsilon$ is the
Clausius-Mossotti relation.  This relation can be
easily generalized to an anisotropic medium by replacing the
scalar dielectric function and other scalar variables by a
dielectric tensor and other tensor variables~\cite{levy}.  Thus,
we can generalize the lattice dispersion relation~\cite{dda1},
used to connect $\alpha$ to $\epsilon$ in the isotropic case, to
the anisotropic case.   To obtain the position-dependent
anisotropic polarizability tensor $\tilde{\alpha}$, we insert the
position-dependent anisotropic dielectric tensor into this
generalized lattice dispersion relation.

To apply this approach to the liquid crystal morphologies discussed above, we
need to transform the local dielectric tensor from a frame of
 reference in which it is diagonal (the ``diagonal frame")
into the lab frame, using a
suitable similarity transformation.  In practice, this
transformation is not difficult, since the local $\tilde{\epsilon}$ and
$\tilde{\alpha}$ are simultaneously diagonalizable.  In practice, we find
that to calculate a given diagonal element of the lattice dispersion
relation
polarizability, we only need the same diagonal element of the
dielectric tensor.  Thus, the diagonal polarizability
tensor using lattice dispersion relation can be calculated in this diagonal
frame.
Also, the lattice dispersion relation is invariant under the above
similarity transformation.  Thus, if the dielectric tensor is initially
diagonal in some frame of reference, we can obtain the diagonal
polarizability using lattice dispersion relation in this frame first, then
carry out a suitable
similarity transformation on this diagonal  polarizability to get
the polarizability in the lab frame.

To check the convergence of our DDA calculation, we considered
an liquid-crystal-coated
nanoparticle with a gold radius of 30nm, surrounded by an nematic
liquid crystal of 30 nm thickness. We carried out the DDA
calculations with a variable number of point
objects representing the liquid-crystal-coated gold nanoparticle.
This was accomplished by varying the ``linear mesh size" $M \equiv
2R/d$, where $d$ is the distance between adjacent point objects on a
simple cubic mesh, and $R$ is the radius of a metal nanoparticle,
i.\ e.\ 30nm in this case. In Fig.\ 2(a), we plot the scattering
cross section as a function of this linear mesh size. As can be
seen, the peak position converges well with increasing linear mesh
size $M$, but its height converges less well. We observe the same
behavior in the case of a nanoparticle with Drude dielectric
constant without an liquid crystal coating layer embedded in a
homogeneous medium as shown in Fig 2(b). Compared with the exact Mie
theory, the peak position converges well even in case of $M=20$,
while its height converges more slowly (within about 5 \% in case of
$M=120$). In Fig. 2, we use Drude dielectric function with
$\omega_P\tau=7.7$. Thus, our DDA results for the {\em
position} of the surface plasmon peak from an liquid-crystal-coated
nanoparticle are robust against changes of the linear mesh size,
even if we consider scattering from position-dependent anisotropic
media. Hence, we can calculate the positions of the scattering peaks
with and without an liquid crystal coating, and for various coating
thicknesses, using the DDA, and compare
the results to measurements of Ref.\ \cite{muller}.

As a first calculation with the actual gold dielectric function,
we studied how the surface plasmon peak position changes as a function of
liquid crystal
layer thickness, under the assumption that the director field
is uniformly oriented in one direction parallel to the substrate.
Here for the dielectric function of a gold nanoparticle, we used a
finite-particle-size corrected dielectric function, which we can
estimate from the tabulated values of bulk gold~\cite{gold}. In
this case, we calculated the scattering cross section for an
incident electric field polarized parallel to the liquid crystal director,
i.\ e. $\theta=0^{o}$ in Fig. 1(b). By comparing the position of
this calculated surface plasmon peak to that measured in Ref.\ \cite{muller},
using a similar setup, we estimate that the thickness of the liquid crystal
layer in those measurements is $\sim 30$ nm, comparable to the
radius of the gold particle.

In Fig.\ 3, we plot the position of the plasmon peak versus
polarization angle of the incident light, for three liquid
crystal coating morphologies.  Two of the morphologies are the
``boojum pair" and ``baseball" configurations described above, and
the third is a director uniformly oriented in one direction
parallel to the substrate.   A morphology with two singular points
of the north-south pole type shows the biggest peak shift (about
0.023 eV) with change in polarization angle; the magnitude of this
shift makes the best fit to the value ($\sim 0.030$ eV) observed
in the experiments.   By contrast, the value obtained when we
assume a uniform director field is about $0.011$ eV, and that
obtained for the baseball-like geometry is the smallest among the
three different morphologies ($\sim 0.004$ eV). Thus, the surface
deformation of the director field, which is caused by the gold
nanoparticle surface and by the boundaries with host medium or the
substrate, can strongly influence and even enhance the plasmon
splitting of this system. This enhancement suggests that the
optical properties of such nanoscale systems can be controlled
electrically, by inclusion of liquid crystals.

The present calculations suggest that the liquid crystal film in the
experiments~\cite{muller} is probably relatively thick ($\sim 30$
nm).  If the coating were thinner, the stronger surface perturbation
might lead to a baseball-like configuration as suggested in Ref.\
\cite{nelson}, and hence, a much smaller surface plasmon splitting than seen in
the experiments.   We speculate that there may be a boojum pair-baseball
crossover as the film gets thinner.  If so, monitoring the splitting
as a function of liquid crystal coating thickness would provide an optical
signature of such a crossover.  Moreover, if there is a strong
suppression of the plasmon splitting, as we find in case of a
baseball-like configuration, this would be an experimental indication
that the coating thickness is suitable for forming tetravalent
functionalized points around the gold nanoparticles, which can
realize a tetravalent chemistry of colloids~\cite{nelson}.

Finally, we briefly discuss a liquid-crystal-coated gold nanoparticle in an
applied electric field. If we assume the tangential boundary condition,
two boojum defects are induced on the surface of the
nanoparticle~\cite{lubensky}.  In this case, the morphology of liquid crystal
director near the surface of the gold nanoparticle, will be the same as the
boojum pair case
in zero applied electric field, and thus we expect that the splitting of
the peak in the scattering cross section should have the same
enhancement by surface deformation as in zero applied field.  The
effects of an electric field, and more general boundary conditions, may need
further investigation.

The present results may have a wide range of applications in
nanoscience and nanotechnology, because they show that optical
properties of systems containing nanoparticles can be sensitively
controlled with the use of liquid crystals.  The transmission and
absorption of such materials could be tuned by a dc electric field,
which will alter the liquid crystal director field, or by
controlling the surface interactions between metal and liquid
crystal.  This opens up the possibility of new classes of tuneable
electro-optic materials based on nanoscale metallic particles
combined with liquid crystals.  Moreover, the desirable boojum pair
configuration may form by self-organization, thereby allowing new
types of self-assembled nano-optical devices. Finally, a tetrahedral
baseball-like director configuration is a potential building block for
a self-assembled three-dimensional photonic crystal with a diamond
structure.  The measured surface plasmon splitting may be a valuable
diagnostic to determine when this baseball configuration forms.

This work was supported by NSF Grant DMR04-13395; calculations were
carried out, in part, using the Ohio Supercomputer Center. SYP
thanks Professor D. R. Nelson for a useful discussion and for
permission to reproduce Figs. 1(c) and 1(d).

\newpage
\begin{figure}
\scalebox{0.5}{\includegraphics{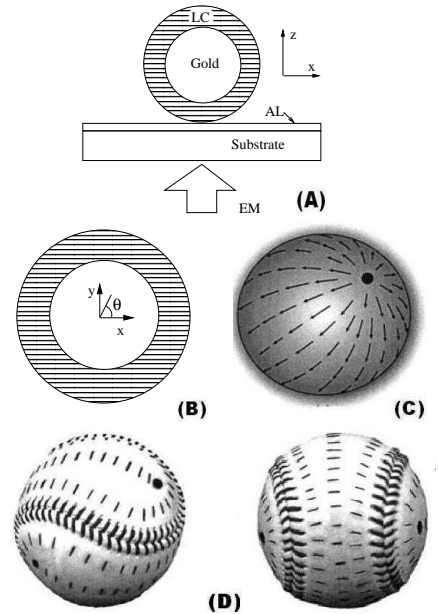}}

\caption{(a) Schematic of the experimental geometry considered in
these calculations. We assume that a spherically coated nanoparticle
sits on a substrate.  The substrate has a thin coating which aligns
the liquid crystal director at the point of contact in the $x$
direction. Monochromatic linearly polarized light (EM) is assumed
normally incident as shown.  The three possible liquid crystal
morphologies which we consider are shown schematically as (b)
homogeneous; (c) ``boojum pair"; and (d) ``baseball." We calculate
the position of the surface plasmon peak as a function of the
polarization angle $\theta$ indicated in Fig.\ 1(b). }
\end{figure}

\begin{figure}
\scalebox{0.55}{\includegraphics{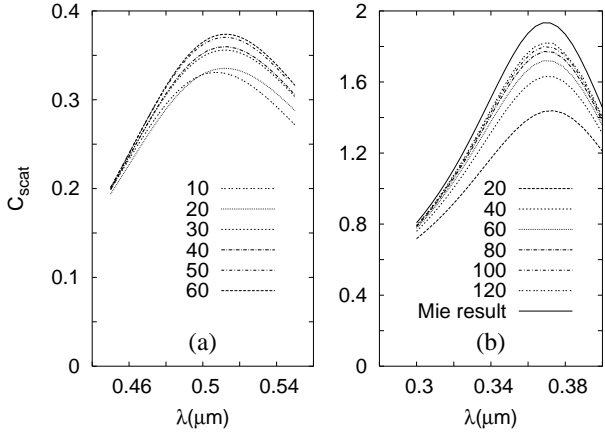}}
\caption{ Calculated scattering cross section versus wavelength
$\lambda$ for (a) an liquid-crystal-coated metal nanoparticle at
angle $\theta = 0^{o}$ for boojum pair configuration with radius
30nm, liquid crystal coating thickness 30nm, and varying the
DDA mesh size, as indicated in the legend;
(b) same for a bare metal nanoparticle and a radius 30nm in an air
host with various mesh sizes. Solid line indicates the exact Mie
result. Here we used Drude dielectric function, for simplicity.}
\end{figure}

\begin{figure}
\scalebox{0.65}{\includegraphics{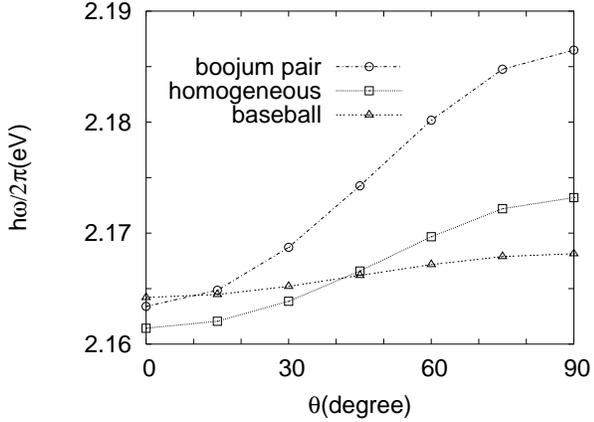}}

\caption{Frequency of peak in scattering cross section for an
liquid-crystal-coated gold nanoparticle (with the actual gold dielectric
function with finite-particle-size correction) versus angle
$\theta$, for boojum pair, homogeneous, and baseball
configurations. }
\end{figure}

\end{document}